\documentclass[a4paper,11pt]{article}
\usepackage{pos}

\title{An Exotic Path to Glue states Decay}
\ShortTitle{Exotics and glueballs}

\author*[a]{Jean-Marie Fr\`{e}re  }

\affiliation[a]{Theoretical Physics CP225, Universit\'{e} Libre de Bruxelles,\\
  blvd du Triomphe , 1050 Brussels, Belgium, \\ and Brout-Englert-Lema\^itre Center, Brussels}

\emailAdd{frere@ulb.be}

\abstract{Glueballs are the most obvious, but this far not fully tested prediction of Quantum Chromodynamics.
The lowest expected glueball states, in particular $0^{++}$ states are difficult to characterize, since they share the quantum numbers of the "vacuum" and of a number of possible quark-antiquark states. In this paper, we argue that the more easily identifiable exotics provide us with strong hints on the way to pursue the chase, and in particular the relation with quantum anomalies and the $\eta$ and $\eta '$ states.}

\FullConference{Corfu Summer Institute 2023 "School and Workshops on Elementary Particle Physics and Gravity" (CORFU2023)\\
 23 April - 6 May , and 27 August - 1 October, 2023\\
Corfu, Greece\\}

\begin{document}
\maketitle

\section{Introduction to a difficult hunt}
The most straightforward prediction of Quantum Chromodynamics (QCD) is obviously the possibility of colour-neutral (and hence unconfined) states made solely of gluons.
While their precise characterization in terms of composite states is difficult (gluons don't exist "on shell", and hence an approach similar to the building blocks of the quark model is difficult), they have been studied by lattice QCD. Typically, the lowest states are thought to be "2-gluon" singlets, with possible quantum numbers $0^{++}$ and $2^{++}$.

A number of states with $0^{++}$ (spin 0, and flavour-less)  (in the current classification they are named $f_0$ ) have been observed over time in different production schemes and in various decay channels. It proves however very difficult to distinguish them from "ordinary" quark states.
Since they are relatively broad states, one may also fear that phase space distortion may lead to their classification splitting into  different states depending on the decay observed, pending a full coupled channel analysis.

Only some very rough  (and somewhat dubitable) criteria exist for selecting more specific decays.
It is correctly advocated that the decay should be "flavour-blind" (as gluons carry no flavour), but this only holds at the level of the  interaction Lagrangian, and  even limiting oneself to 2-body decays, other arguments states that "overlap of wave functions" (and the glueball candidates are heavier than 1 GeV) would favour the heavier mesons (think $K$ pairs rather than $\pi$ pairs. Another line of thought is also brought in to favours heavier quarks, observing that decay diagrams into pseudoscalars would suffer from chiral suppression. \textbf{This last argument however falls through }when dealing with anomaly graphs (involving for instance the $\eta$ and $\eta$' states.

This also does not take into account multi-particle decays. For instance, we can argue that lighter $f_0$ states could be favoured (in that they can "mix" with glue states also sharing the vacuum quantum numbers). One such possibility has already been seen for the $f_0(1500)$  with the final state.
$$f_0 \rightarrow \sigma \sigma \rightarrow ( 2 \pi ) ( 2 \pi )$$ where $\sigma$ (or in current nomenclature $f_0(500)$ , the $0^+$ partner of the pion itself a relatively elusive state only seen by careful multichannel analysis.

Another (and in our view more promising) way to characterize the glue-involving states is to consider the $\eta$ and $\eta '$ final states, both related to 2-gluon states by the quantum anomalies
$$\partial^\mu A_\mu^0 = \frac{2}{\sqrt{3} }(m_u \overline{u} i \gamma_5 u + m_d \overline{d} i \gamma_5 d + m_s \overline{s} i \gamma_5 s)+ \frac{1}{\sqrt{3} } \frac{3 \alpha_s}{4 \pi} G^{\mu \nu} \widetilde{G_{\mu \nu}}$$.
This approach was suggested early on by Gherstein et al.\cite{Gherstein} , and later developed by us \cite{Akhoury:1987ed} \cite{Ball:1995zv} \cite{Frere:2015xxa} \cite{Escribano:2005qq}.

\section{Hybrid states could provide the key, ... but we were misled! }
Hybrid states are typically composed of quarks (in particular a quark-antiquark pair) and at leat one gluon. As such, they may escape the usual selection rules of the ordinary mesons (in terms of accessible $J^{P,C}$ states), in which case they will be called "exotics". Such characteristics should make them conspicuous in the forest of ordinary mesons.

Obviously, observing a "wrong J, P, C " state is still somewhat ambiguous, as the quantum numbers of a gluon can be mimicked by a quark pair, like $ \overline{\psi} \gamma^\mu \lambda_a \psi$ , similar to $G^\mu_a$ : this means that a (quark-antiquark-gluon) exotic could be mistaken with a "4-quark" state.

One such state was claimed early \cite{IHEP-Brussels-LosAlamos-AnnecyLAPP:1988iqi}, with quantum numbers similar to the $\rho$ but opposite paritly $J^{PC} = 1^{-+}$ instead of $J^{PC} = 1^{--}$, with same isospin. In current notation, it would be one of the $\pi_1$ states.
It turned out that it took a long time and several experiments to become accepted in the mainstream!
We could think of such a state as associated to the interpollating field

$$\phi^\mu_{\pi_1} \equiv g_s  G^{\mu \nu}_a  \frac{1}{2} ( \overline{u }\gamma_\nu \gamma_5 \lambda^a u) \frac{1}{f_{\pi_1}m_{\pi_1}^3} )$$

It turned out that it took a long time and several experiments to become accepted in the mainstream! One of the main reasons was the observed decay mode. At the time  theoretical prejudice was that $\eta \pi^0$  would be disfavoured with respect to a dominant $\rho \pi$ decay.
This can be traced back to a paper \cite{DeViron:1984svx}, which, while ground-breaking in many respects, unfortunately did not include the possible contribution of the quantum (Steinberger-Adler-Bell-Jackiw) anomaly.

\textbf{This result - $\rho \pi$ decay favored over $\eta(') \pi)$ - went unquestioned for a long time, despite growing experimental evidence and our later calculation \cite{Frere:1988ac}, which strongly suggested the opposite.}

Let us try to summarize the current situation with the $I=1, J^{PC} = 1^{-+}$ states.

Now, several collaborations have "seen" such states (COMPASS, Crystal Barrel,...). Two main entries exist in Particle Data Table, namely $\pi_1 (1400)$ and $\pi_1 (1600)$.
The first is consistently seen in $\eta \pi$ and the second in $\eta' \pi$ while other
decay modes are less consistently observed  ($b_1 \pi$, $\rho \pi$).

A re-analysis of the COMPASS data by the JPAC group using a multichannel analysis has come to an interesting result, as they best describe the situation with one single (distorted) state \cite{JPAC:2018zyd}.
This analysis is confirmed  by Kopf  et al. \cite{Kopf:2020yoa} using in addition the Crystal Barrel data .
They even extract a ration (see the paper for the systematic errors)

 $$\Gamma (\pi \eta') / \Gamma (\pi \eta) = 5.54 \pm 1.1 (stat)$$
This  ratio is large despite phase space suppression for the $\eta'$
Note that this ratio is similar to the ratio $$ \frac{J/\psi \rightarrow \gamma \eta'}{J/\psi \rightarrow \gamma \eta}$$

This time, the ratio is again large despite the fact that the $\eta'$ mode is affected by a p-wave suppression: it  was the basis for Gherstein's suggestion of a connection between anomalies, glueballs and the $\eta$ family.

Unfortunately, we cannot extract the matrix element ratio  for the $\pi_1$ branchings into $\eta(')$ channels, since the phase space  of the stretched  $\pi_1$ is anything from simple.

In contrast, the more "popular" decay mode $\pi_1 \rightarrow \rho \pi$ seems much more difficult to isolate. A fit (using 88 partial waves!) has recently isolated it in the COMPASS data \cite{COMPASS:2021ogp}, this article also discusses the previous attempts in other experiments.

\begin{figure}[h]
\begin{center}
\includegraphics [width=9cm]{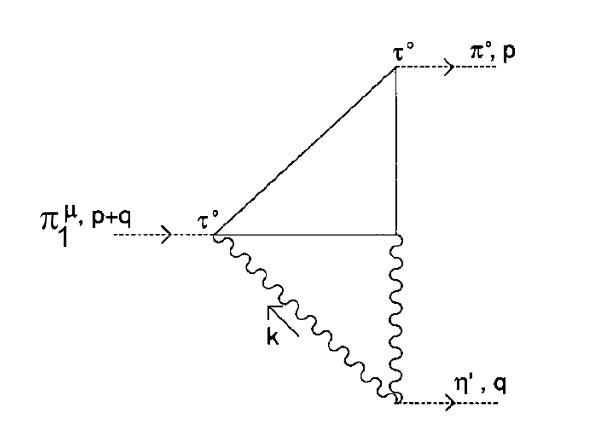}
\caption{A mechanism for the exotic $\pi_1$ decay through the anomaly}
\label{titard}
\end{center}
\end{figure}

This gives support to our hypothesis in \cite{Frere:1988ac} that the decay is similarly  largely mediated by a 2-gluon process, and hence the $\eta' to \eta$ ratio receives a contribution proportional to (see fig.1)
$$ | \frac{<|G^{\mu \nu} \widetilde{G_{\mu \nu}}| \eta' >}{<|G^{\mu \nu} \widetilde{G_{\mu \nu}}| \eta >} |^2 $$

\begin{figure}[h]
\begin{center}
\includegraphics [width=10cm]{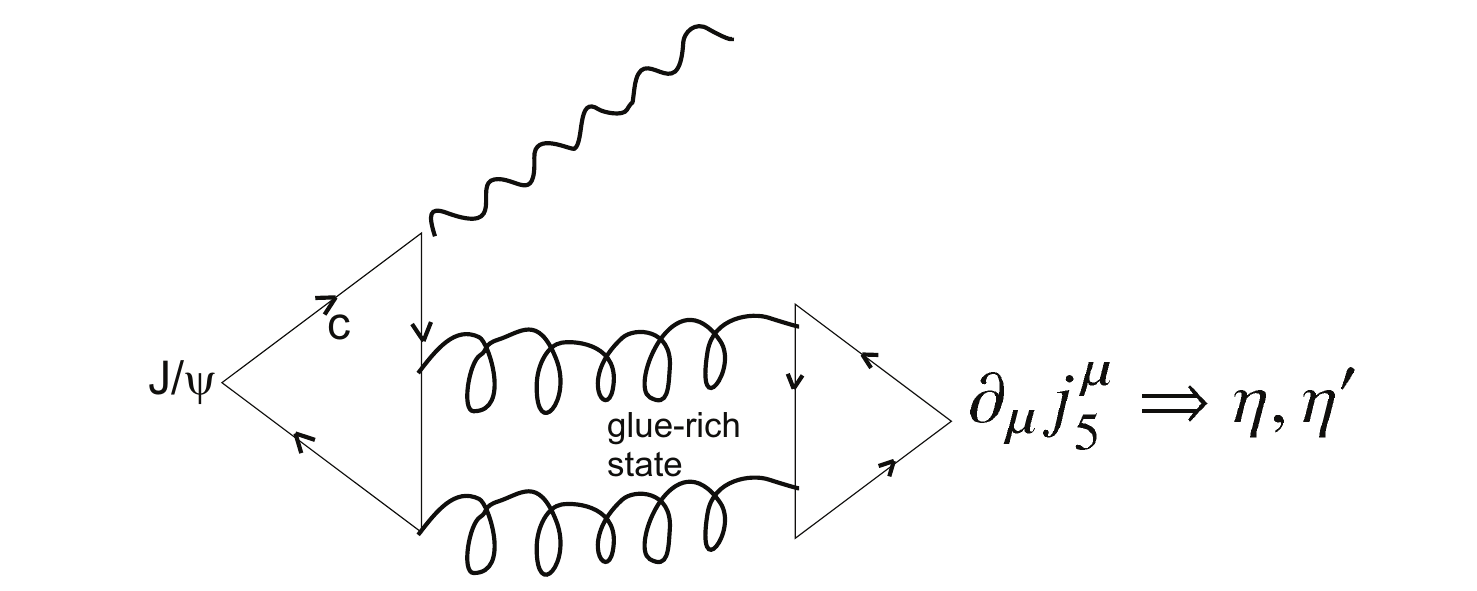}
\caption{J/$\psi$ radiative decay as a glue-rich source coupled to the $\eta$('); notice that no mass insertion is needed in the right-most triangle, therefore negating the "chiral suppression"}
\label{JradDecay}
\end{center}
\end{figure}

We will not try to be more quantitative here (while we can extract the ratio of the matrix elements from the $J/\psi$ decay, we already stated that the phase space is difficult to evaluate in the present case due to the distorted shape of the $\pi_1$ wave).

\section{A clean gluon source and $\eta_1$ (and a further quest for $\pi_1$).}
Since the characterization of glueballs and hybrids is not easy on the basis of their sole decays, it is advantageous to look in already "glue-rich" production channels, as we have previously argued \cite{Frere:2023mxb} ; this was also an argument we made in favor of the COMPASS set-up.
In the previous paragraph, such could be found in central production, but an even cleaner signal is found in radiative decays of quark-antiquark mesons, as can be seen in fig. \ref{JradDecay}.

We are not aware of a search for $\pi_1$ in this context (as it is probably suppressed by isospin), but a similar state, $\eta_1$ has been reported by the BESIII collaboration \cite{BESIII:2022riz} \cite{BESIII:2022iwi}.
Quite interestingly it was observed through its $\eta \eta'$ decay (analogous to the previous $\eta \pi$
 transposed to the I=0 channel). Independently of our much older publication,
  a similar mechanism for the decay into $\eta \eta'$ was proposed
  by ref. \cite{Chen:2022qpd} \cite{Chen:2022asf}.

This validates our approach, and in this first (exotics-centered) step highlights the crucial importance of the axial anomaly in relating the gluon states to the light neutral mesons $\eta$ or $\eta'$.
Not only are the respective decays $\pi_1 \rightarrow \eta' \pi$ and $\eta_1 \rightarrow \eta' \eta$ allowed, they are indeed dominant, as we expected a long time ago!
The prejudice that  $\pi_1 \rightarrow \rho \pi$, \emph{\textbf{based on a calculation which did not take into account quantum anomalies should thus be discarded}} !

As already mentioned, isospin suppression does not suggest the radiative $J/\psi$ decays as an equally clean source for the study of the $\pi_1$! Yet, we note that BESIII \cite{BESIII:2016gkg} has observed $J/\psi \rightarrow \gamma \pi^0 \eta$, where a more extensive partial wave analysis could be of interest. (the Particle Data Group does not quote a similar entry for the $\eta'$ counterpart).

\section{What does this teach us for the glueball states? }

The above sections have mainly dealt with the exotics (hybrids), where quantum numbers help in identifying the presumably valence-gluon-containing states (with the ambiguity of possible 4-quark states mimicking them).

We want now to return to the possibly more fundamental question of glueballs (unconfined colour neutral states made from gluons alone), which are a direct prediction of pure QCD independently of quarks.

In a way, we are confronting two worlds here : one composed of gluon bound states, which have no direct link to ordinary mesons, and the other made of the (meson containing) usual mesons, which are the only possible decay channels (since the gluons don't have electroweak interactions).
We have previously advocated that the quantum anomalies could be the bridge between those two worlds.

It is thus conceivable that the $0^{++}$ and $0^{-+}$ combinations - scalars like the $\sigma$ and pseudoscalars like the $\eta(')$-, which share quantum numbers respectively with the glue-only combinations $ G^{\mu \nu}  G_{\mu \nu} $ and
$ G^{\mu \nu} \widetilde{ G}_{\mu \nu} $ respectively play an important role in the decay channels.

Glueball candidates have been proposed for a long time (we will return shortly to the remarkable history of the GAMS collaboration\cite{Serpukhov-Brussels-AnnecyLAPP:1983xdr} \cite{Serpukhov-Brussels-AnnecyLAPP:1983jxn}, and gluon-rich production can be expected in central collisions, but also in the "clean" $J/\psi$ radiative decays.

For the time being, a series of such candidates ($f_0(xx)$) is found in Particle Data \cite{PDG2022}, and it is difficult to distinguish which is predominantly a glueball.
It may also be possible that a similar scenario to the one observed for the $\pi_1$ occurs, meaning that a more extensive coupled channel analysis may bring light on related states, as we are in a
similar situation of heavily distorted phase space.
An attempt at such an analysis can be found in \cite{Sarantsev:2021ein}, but it does not seem to include the $J/\psi \rightarrow \gamma \eta \eta'$ modes, which will prove essential.

Some time ago, we have followed a popular approach of exploiting the topologies of various decay channels
\cite{Frere:2015xxa} to try to sort out those modes and tried to include anomaly related contributions. It must be stated that such an approach is not a rigorous (Feynman diagrams-like) calculation, as we don't know the exact wave functions and momentum dependence of the channels. The situation however remains muddy, and expectations don't seem to be fully reproduced (at least in the current analysis of the experiments).
\begin{figure}[h]
\begin{center}
\includegraphics [width=12cm]{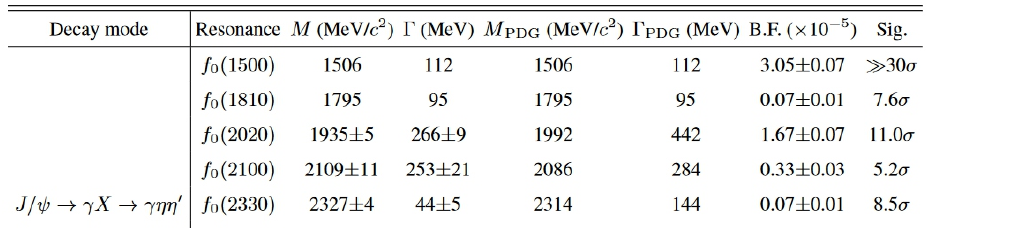}
\caption{possible glueballs in radiative  $J/\psi$ decay as a glue-rich source coupled to the $\eta \eta'$, The product of the branchings is a possible indicator of glueballs; this is part of a table from \cite{BESIII:2022iwi} }
\label{BESiiiTable}
\end{center}
\end{figure}

Instead, we find that  BESIII provides us with a particularly interesting approach,\textbf{ the combined branching ratio} $J/\psi \rightarrow \gamma X , X \rightarrow final state$ \textbf{which effectively quantifies the product of "glue-rich production" and of "glue-rich decay".} The table of fig. \ref{BESiiiTable} is extracted from their paper. \cite{BESIII:2022iwi}

We see indeed that (we refer here to the PDG labeling of the states, each of which appears at different masses in different experimental settings) 2 states below 2GeV   are salient, namely the $f_0(1500)$ and $f_0(1810)$ (we will relate them with the first observations by GAMS in the next section).

\textbf{{In particular, and despite the strong phase-space suppression, the $f_0(1500)$ is very prominent in the combined branching ratio.} }

\vspace{5mm}

The $f_0 (1500)$ is actually quite a strange "beast" if we envisage it as a quark state; as can be seen from its decay mode (see table \ref{Gdecaymodes}.
\begin{table} [h]
\centering
\begin{tabular}{|l|r|}
  \hline
  $f_0(15000) $ & decay fraction ( \%) \\
    \hline
  $\pi \pi$  & 34.5 $\pm$ 2.2 \\
 $ 4 \pi $& 48.9 $\pm$ 3.3 \\
  $ \eta \eta$ & 6.0 $\pm$ 0.9 \\
  $ \eta \eta'$ & 2.2 $\pm$ 0.8 \\
  $ K \overline{K} $ & 8.5 $\pm$ 1.0 \\
  $ 2 \gamma $ & not seen  \\
  \hline
\end{tabular}
\caption{$f_0(1500)$ branching ratios according to PDG \cite{PDG2022}}
\label{Gdecaymodes}
\end{table}

As seen here, it decays predominantly to $ 4 \pi$ , which is evocative of a possible intermediary state of ($\sigma$ stands for $f_0 (500)$ )  $$ f_0 \rightarrow 2 \sigma \rightarrow (2 \pi + 2 \pi ) $$

Quite remarkably also, the $\eta \eta'$ mode is large as we would expect from a large (gluon-anomaly) contribution ; (let us keep in mind that it is very strongly suppressed by phase space - a detailed coupled channel simulation would be needed to extract the matrix elements).

\section{A short tribute to the GAMS pioneers. }

I must confess that I was brought to the topic of glueballs and exotics through questions from my GAMS colleagues. I had an interest in CP violations and anomalies, and their concern to see the $\eta(')$ so prominent when most of the literature wanted to exclude them started my investigation of the field (which as by-product, led to a unified description of all anomaly related processes, like $P\rightarrow V \gamma$, $V\rightarrow P \gamma$,...)\cite{Ball:1995zv}.

In a succession of papers trough experiments at Serpukhov and CERN, they used a lead-glass wall as the main calorimeter (detection from $\gamma$ gave them a unique approach, and even the insensitivity to the $\rho^0$ peak could be used to an advantage) and of course had particular sensitivity to $\eta(')$.This later led to proposing the  COMPASS experiment.

Among the states first detected were the $O^{++}$ candidates \cite{Serpukhov-Brussels-AnnecyLAPP:1983xdr}\cite{Serpukhov-Brussels-AnnecyLAPP:1983jxn}\cite{Serpukhov-Brussels-LosAlamos-AnnecyLAPP-Pisa:1987bht} \cite{IFVE-Brussels-Annecy-LosAlamos:1991bpt}\cite{IFVE-Brussels-Annecy-LosAlamos:1988ado} \cite{WA102:1999hsn}
, notably what they called the G(1590) (probably the current $f_0(1500)$, X (1750),  X (1920),
close to $f_0 (1770)$, (see however a recent multichannel analysis \cite{Sarantsev:2021ein})

The exotic $1^{-+}$ was in discovered in  \cite{IHEP-Brussels-LosAlamos-AnnecyLAPP:1988iqi} with a "mass" 1405 MeV, now part of the $\pi_1 (1400-1600)$ complex. Once again, a pure prejudice (the neglect of quantum anomalies) against the $\eta \pi$ mode made its recognition difficult. The same holds for their possible X(1740) and X(1910) which could be the current $\eta_1 (1855)$ from BESIII.

There is thus hope now that with the progress in the radiative $J/\psi$ (and possibly even $\Upsilon$ ) decays and in  multichannel analysis, together with the rehabilitation of the $\eta'$ channel too often misunderstood, we could see some clarification in the field of glueballs and exotics, and in so doing, reconcile observation with the most basic QCD prediction.


\begin{thebibliography}{99}

\bibitem {Gherstein}
Gershtein, S.S., Likhoded, A.A. and  Prokoshkin, Y.D. G(1590)-Meson and possible characteristic features of a glueball,
Z. Phys. C - Particles and Fields 24, 385 (1984).

\bibitem{Akhoury:1987ed}
R.~Akhoury and J.~M.~Frere,
``$\eta$, $\eta^\prime$ Mixing and Anomalies,''
Phys. Lett. B \textbf{220} (1989), 258-264
doi:10.1016/0370-2693(89)90048-8

\bibitem{Ball:1995zv}
P.~Ball, J.~M.~Frere and M.~Tytgat,
``Phenomenological evidence for the gluon content of eta and eta-prime,''
Phys. Lett. B \textbf{365} (1996), 367-376
doi:10.1016/0370-2693(95)01287-7
[arXiv:hep-ph/9508359 [hep-ph]].

\bibitem{Frere:2015xxa}
J.~M.~Fr\`ere and J.~Heeck,
``Scalar glueballs: Constraints from the decays into $\eta$ or $\eta'$,''
Phys. Rev. D \textbf{92} (2015) no.11, 114035
doi:10.1103/PhysRevD.92.114035
[arXiv:1506.04766 [hep-ph]].

\bibitem{Escribano:2005qq}
R.~Escribano and J.~M.~Frere,
``Study of the eta - eta-prime system in the two mixing angle scheme,''
JHEP \textbf{06} (2005), 029
doi:10.1088/1126-6708/2005/06/029
[arXiv:hep-ph/0501072 [hep-ph]].


\bibitem{IHEP-Brussels-LosAlamos-AnnecyLAPP:1988iqi}
D.~Alde \textit{et al.} [IHEP-Brussels-Los Alamos-Annecy(LAPP)],
``Evidence for a 1-+ Exotic Meson,''
Phys. Lett. B \textbf{205} (1988), 397
doi:10.1016/0370-2693(88)91686-3

\bibitem{DeViron:1984svx}
F.~De Viron and J.~Govaerts,
Phys. Rev. Lett. \textbf{53} (1984), 2207-2210
doi:10.1103/PhysRevLett.53.2207

\bibitem{Frere:1988ac}
J.~M.~Frere and S.~Titard,
``A NEW LOOK AT EXOTIC DECAYS rho-tilde (1-+, I = 1) ---\ensuremath{>} eta-prime pi versus rho pi,''
Phys. Lett. B \textbf{214} (1988), 463-466
doi:10.1016/0370-2693(88)91395-0

\bibitem{JPAC:2018zyd}
A.~Rodas \textit{et al.} [JPAC],
Phys. Rev. Lett. \textbf{122} (2019) no.4, 042002
doi:10.1103/PhysRevLett.122.042002
[arXiv:1810.04171 [hep-ph]].

\bibitem{Kopf:2020yoa}
B.~Kopf, M.~Albrecht, H.~Koch, M.~K\"u\ss{}ner, J.~Pychy, X.~Qin and U.~Wiedner,
``Investigation of the lightest hybrid meson candidate with a coupled-channel analysis of ${{\bar{p}}p}$-, $\pi ^- p$- and ${\pi \pi }$-Data,''
Eur. Phys. J. C \textbf{81} (2021) no.12, 1056
doi:10.1140/epjc/s10052-021-09821-2
[arXiv:2008.11566 [hep-ph]].

\bibitem{COMPASS:2021ogp}
M.~G.~Alexeev \textit{et al.} [COMPASS],
Phys. Rev. D \textbf{105} (2022) no.1, 012005
doi:10.1103/PhysRevD.105.012005
[arXiv:2108.01744 [hep-ex]].


\bibitem{Frere:2023mxb}
J.~M.~Fr\`ere,
PoS \textbf{CORFU2022} (2023), 009
doi:10.22323/1.436.0009
[arXiv:2304.09083 [hep-ph]].


\bibitem{BESIII:2022riz}
M.~Ablikim \textit{et al.} [BESIII],
``Observation of an Isoscalar Resonance with Exotic JPC=1-+ Quantum Numbers in J/\ensuremath{\psi}\textrightarrow{}\ensuremath{\gamma}\ensuremath{\eta}\ensuremath{\eta}',''
Phys. Rev. Lett. \textbf{129} (2022) no.19, 192002
[erratum: Phys. Rev. Lett. \textbf{130} (2023) no.15, 159901]
doi:10.1103/PhysRevLett.129.192002
[arXiv:2202.00621 [hep-ex]].

\bibitem{BESIII:2022iwi}
M.~Ablikim \textit{et al.} [BESIII],
``Partial wave analysis of J/\ensuremath{\psi}\textrightarrow{}\ensuremath{\gamma}\ensuremath{\eta}\ensuremath{\eta}',''
Phys. Rev. D \textbf{106} (2022) no.7, 072012
[erratum: Phys. Rev. D \textbf{107} (2023) no.7, 079901]
doi:10.1103/PhysRevD.106.072012
[arXiv:2202.00623 [hep-ex]].


\bibitem{Chen:2022qpd}
H.~X.~Chen, N.~Su and S.~L.~Zhu,
`QCD Axial Anomaly Enhances the \ensuremath{\eta}\ensuremath{\eta}' Decay of the Hybrid Candidate \ensuremath{\eta} $_{1}$(1855),''
Chin. Phys. Lett. \textbf{39} (2022) no.5, 051201
doi:10.1088/0256-307X/39/5/051201
[arXiv:2202.04918 [hep-ph]].

\bibitem{Chen:2022asf}
H.~X.~Chen, W.~Chen, X.~Liu, Y.~R.~Liu and S.~L.~Zhu,
``An updated review of the new hadron states,''
Rept. Prog. Phys. \textbf{86} (2023) no.2, 026201
doi:10.1088/1361-6633/aca3b6
[arXiv:2204.02649 [hep-ph]].





\bibitem{BESIII:2016gkg}
M.~Ablikim \textit{et al.} [BESIII],
``Observation of $J/\psi\to \gamma\eta\pi^{0}$,''
Phys. Rev. D \textbf{94} (2016) no.7, 072005
doi:10.1103/PhysRevD.94.072005
[arXiv:1608.07393 [hep-ex]].






\bibitem{Serpukhov-Brussels-AnnecyLAPP:1983xdr}
F.~G.~Binon \textit{et al.} [Serpukhov-Brussels-Annecy(LAPP)],
``G (1590): A Scalar Meson Decaying Into Two eta Mesons,''
Nuovo Cim. A \textbf{78} (1983), 313
CERN-EP/83-97.



\bibitem{Serpukhov-Brussels-AnnecyLAPP:1983jxn}
F.~G.~Binon \textit{et al.} [Serpukhov-Brussels-Annecy(LAPP)],
``Observation of Reaction $\pi^- p \to \eta^\prime \eta n$ and a Search for Glueball,''
Sov. J. Nucl. Phys. \textbf{39} (1984), 526
IFVE-83-204.

\bibitem{PDG2022}
R.L. Workman et al. (Particle Data Group), Prog.Theor.Exp.Phys. 2022, 083C01 (2022)
and 2023 update.

\bibitem{Sarantsev:2021ein}
A.~V.~Sarantsev, I.~Denisenko, U.~Thoma and E.~Klempt,
Phys. Lett. B \textbf{816} (2021), 136227
doi:10.1016/j.physletb.2021.136227
[arXiv:2103.09680 [hep-ph]].

\bibitem{Serpukhov-Brussels-LosAlamos-AnnecyLAPP-Pisa:1987bht}
D.~Alde \textit{et al.} [Serpukhov-Brussels-Los Alamos-Annecy(LAPP)-Pisa],
``Production of $G(1590)$ in 300-{GeV} Central $\pi^- N$ Collisions,''
Phys. Lett. B \textbf{201} (1988), 160
doi:10.1016/0370-2693(88)90100-1




\bibitem{IHEP-IISN-KEK-LANL-LAPP:1991nfl}
D.~Alde \textit{et al.} [IHEP-IISN-KEK-LANL-LAPP],
``Further studies of the x (1910) meson,''
Phys. Lett. B \textbf{276} (1992), 375-378
doi:10.1016/0370-2693(92)90334-Z

\bibitem{IFVE-Brussels-Annecy-LosAlamos:1991bpt}
D.~Alde \textit{et al.} [IFVE-Brussels-Annecy-Los Alamos],
``Production mechanism of the X (1740) meson,''
Phys. Lett. B \textbf{284} (1992), 457-460
doi:10.1016/0370-2693(92)90461-C

\bibitem{IFVE-Brussels-Annecy-LosAlamos:1988ado}
D.~Alde \textit{et al.} [IFVE-Brussels-Annecy-Los Alamos],
``Evidence for a 1.9-{GeV} Meson Decaying Into $\eta \eta^\prime$,''
Phys. Lett. B \textbf{216} (1989), 447
doi:10.1016/0370-2693(89)91148-9

\bibitem{WA102:1999hsn}
D.~Barberis \textit{et al.} [WA102],
``A Study of the eta eta-prime and eta-prime eta-prime channels produced in central p p interactions at 450-GeV/c,''
Phys. Lett. B \textbf{471} (2000), 429-434
doi:10.1016/S0370-2693(99)01392-1
[arXiv:hep-ex/9911041 [hep-ex]].












\end{thebibliography}
\end{document}